\documentclass{article}
\usepackage{textgreek}
\usepackage{url}
\usepackage{mathtools}
\usepackage{graphicx}
\usepackage{cite}

\begin{document}

\title{A Simple Linear Model to Aid in Analyses\\
of the \textbeta{} Pictoris Moving Group}

\author{
Josina~O.~do~Nascimento\\
Observat\'orio Nacional\\
Rua General Jos\'e Cristino, 77\\
20921-400 Rio de Janeiro - RJ, Brazil\\
\\
Valmir~C.~Barbosa\thanks{Corresponding author (valmir@cos.ufrj.br).}\\
Programa de Engenharia de Sistemas e Computa\c c\~ao, COPPE\\
Universidade Federal do Rio de Janeiro\\
Centro de Tecnologia, Sala H-319\\
21941-914 Rio de Janeiro - RJ, Brazil}

\date{}

\maketitle

\begin{abstract}
We build a four-dimensional linear model of object membership in the \textbeta{}
Pictoris moving group (BPMG), using two nested applications of Principal
Component Analysis (PCA) to high-quality data on about 1.5 million objects.
These data contain the objects' galactic space velocities and also their
\emph{Gaia} $G$ magnitudes. Through PCA, they ultimately result in a
four-dimensional straight line, referred to as PC~$1'$, about which both the
bona fide members used to obtain the straight line and the candidate members
used to test the model congregate at generally small distances. Our bona fide
members come from a recent, \emph{Gaia} DR2-based compilation. Most candidate
members are from a compilation from 2017. Using a standard procedure to flag
groups of outliers in data sets, we argue that flagging the few possible
outliers we identified on account of distances to PC~$1'$ is consistent with the
nature of the candidate members in use. The spatial and kinematic measurements
that backed their inclusion in the 2017 compilation were of course from before
the availability of data from the \emph{Gaia} mission. Moreover, their radial
velocities at the time were either unknown or estimated somewhat unreliably. We
propose that PC~$1'$ be added to the tool set for BPMG analyses and potentially
extended to other young stellar moving groups.

\end{abstract}

\newpage
\section{Introduction}
\label{intro}

Of all the young stellar moving groups discovered and studied to date, the
\textbeta{} Pictoris moving group (BPMG) \cite{zsbw01} seems to offer the most
tantalizing possibilities for research on planet formation and exoplanets. Not
only is it closest to Earth, but also the star that gives BPMG its name has two
confirmed planets orbiting it \cite{lbcaebgrmlk10,l-etal19}. These
characteristics have motivated substantial effort in the last two to three
decades to discover the group's full suite of members. To the best of our
knowledge, the most complete assessment of the BPMG membership dating from
before the availability of data from the \emph{Gaia} mission is to be found in
\cite{saklf17}. This was soon followed by further comprehensive studies
(cf., e.g., \cite{gmmrrlfrprd18,ls19}), with \cite{gmmrrlfrprd18} already
incorporating data from \emph{Gaia} DR1 and subsequently undergoing updates as
further data releases were made available.

The research underpinning studies such as these is based on spatial, kinematic,
photometric, and several other types of measurement, the kinematic part
comprising the usual galactic space velocities, viz., $U$, $V$, and $W$. The
focus on these velocities has been  widespread since relatively early on
(cf., e.g., \cite{zs04,tqsrms06,tqms08}), since a moving group's members tend to
cluster together in three-dimensional velocity space. This is always important
information, something to be taken into account when looking to confirm new
members, though checking the membership status of a new candidate based on such
clusters must necessarily grapple with criteria from cluster analysis that tend
to be ill defined \cite{e02}.

Here we demonstrate that, by bringing in a fourth dimension to the data space,
a new perspective is gained that can assist in the task of determining cluster
membership by essentially requiring distances from points to a straight line in
four-dimensional space to be calculated. With these distances available for all
objects of interest (bona fide group members as well as candidate members),
outliers can be flagged for further consideration on the basis of how such
distances are distributed. Though in our initial trials with the data we
observed this strategy to make sense for more than one choice of fourth
dimension and more than one moving group, in this paper we deal exclusively with
the fourth dimension being given by the \emph{Gaia} $G$ magnitude \cite{w18} and
with the BPMG.

The possibility of running a straight line through such a four-dimensional
representation of a proposed BPMG membership becomes evident as one examines any
of the three two-dimensional plots involving dimension $G$ ($G\times U$, etc.).
However, pinning the line down through some generalization of simple linear
regression to more than two dimensions would be fraught with difficult
decisions, including deciding which variables to take as independent, something
that already plagues the two-dimensional case in several domains. Our analysis
in this work takes a longer, though still simple, route.

Aiming to make the representation of the data independent of measurement units
and also amenable to visualization in a greater number of two-dimensional plots
(six rather than three), we work on standardized versions of $U,V,W,G$ (i.e.,
versions whose means equal $0$ and standard deviations equal $1$) and moreover
rotate the resulting data in such a way that they become represented by
uncorrelated coordinates. Such rotation results from the well-known method of
Principal Component Analysis (PCA) \cite{jc16}, whose output is essentially the
appropriate $4\times 4$ rotation matrix, henceforth denoted by $R$. Crucially,
$R$ must at bottom function as a model of how points of coordinates $U,V,W,G$
should be transformed to acquire uncorrelated coordinates. As such, it must take
into account as much high-quality data as can be obtained. This holds also for
the desired four-dimensional straight line, which can be obtained by a further,
restricted use of PCA.

We proceed by first describing how we acquired and processed all the necessary
data in Section~\ref{dap}, then presenting results in Section~\ref{res}.
Concluding remarks follow in Section~\ref{concl}.

\section{Data acquisition and processing}
\label{dap}

We focus on the accounts of BPMG membership given in \cite{saklf17} and in
\cite{chbpgppvgs21}. The former of these predates the availability of data from
the \emph{Gaia} mission. The latter, in turn, is based on an update of the
methodology of \cite{gmmrrlfrprd18} to contemplate data from \emph{Gaia} DR2.
Owing to this, and to the selection criterion used in
\cite{chbpgppvgs21}---only objects to which the model in \cite{gmmrrlfrprd18}
ascribes a probability greater than 0.9 are admitted---our set of bona fide
members is taken from the resulting list.

At a certain point in \cite{saklf17}, the search for BPMG members narrows down
to a list of 104 candidate members. Of these, 26 were members already known
from the literature, 41 turned out to be new members, and 37 were rejected. The
26 known members appear in a final list, along with additional members also
known from the literature that did not make the candidate list, amounting to a
total of 146. The 41 new members constitute a list of their own. Our versions of
these two lists are given in Tables~\ref{bk} and~\ref{bn}, containing
respectively the 146 known members and the 41 new members. These two tables
contain information obtained during our own process of data acquisition.

The BPMG membership accounted for in \cite{chbpgppvgs21} contains 64 objects, of
which 48 already appear in Table~\ref{bk}, 5 in Table~\ref{bn}, and the
remaining 11 are given separately in Table~\ref{bx}. Those already in
Table~\ref{bk} or Table~\ref{bn} are marked with asterisks. Whenever needed, we
henceforth refer to the corresponding sub-tables as Table~\ref{bk}(*) and
Table~\ref{bn}(*).

\paragraph{Data acquisition.}
All our data were acquired from the SIMBAD astronomical database
(\url{http://simbad.u-strasbg.fr}) in two separate downloads, both on March 29,
2022, with minor adjustments occurring in the following weeks. Our data,
therefore, already contain those that \emph{Gaia} EDR3 made available.

The first download targeted the 198 objects in Tables~\ref{bk}--\ref{bx},
retrieving for each one: right ascension and declination ($\alpha,\delta$) and
their quality, proper motion in right ascension and in declination
($\mu_\alpha,\mu_\delta$) and their quality, parallax ($\pi$) and its quality,
radial velocity ($\rho$) and its quality, $G$ and its quality. Each quality is
either one of A through E (highest to lowest) or -, the latter indicating that
the corresponding data are missing. A tally of this download is shown in
Table~\ref{tab1} in regard to data completeness and quality (AAAAC indicates
that the spatial and kinematic data are all of quality A while $G$ is of quality
C, which incidentally is the only quality we ever obtained for $G$, not only in
this download but in the second one as well). Note that all 36 objects in this
download for which data were found to be incomplete serve no purpose in this
work. Only the remaining 162 matter.

\begin{table}[t]
\caption{BPMG membership as considered in \cite{saklf17} (known members,
cf.\ Table~\ref{bk}; new members, cf.\ Table~\ref{bn}) or in \cite{chbpgppvgs21}
but not already in \cite{saklf17}, cf.\ Table~\ref{bx}.}
\label{tab1}
\centering
\makebox[\textwidth][c]{\begin{tabular}{lcccc}
\hline
Data availability & &  & & \\
and quality & Table~\ref{bk} & Table~\ref{bn} & Table~\ref{bx} & Total\\
\hline
Complete data, AAAAC & \hskip0.5em 81 & 18 & \hskip0.5em 5 & 104 \\
Complete data, others & \hskip0.5em 32 & 20 & \hskip0.5em 6 & \hskip0.5em 58 \\
Complete data (total) & 113 & 38 & 11 & 162 \\
Incomplete data & \hskip0.5em 33 & \hskip0.5em 3 & \hskip0.5em 0 & \hskip0.5em 36 \\
Grand total & 146 & 41 & 11 & 198 \\
\hline
\end{tabular}
}
\end{table}

The second download expanded on the first by targeting a considerably larger
body of objects, aiming to put together as many quality-AAAAC objects of
appropriate type as available in the database. As such, it necessarily retrieved
the 104 quality-AAAAC objects already retrieved in the first download, so
duplicates had to be discarded when putting together the quality-AAAAC portion
of the final data set. As anticipated in Section~\ref{intro}, the intended use
of the data in this second download was to feed the determination of matrix $R$.
The following SIMBAD object classification types were used in the queries:
10.12.00.00 (Possible peculiar star) and subtypes,
12.13.00.00 (Double or multiple star) and subtypes,
14.00.00.00 (Star) and subtypes. (In SIMBAD query notation, these can all be
retrieved by specifying \verb|maintypes = 12.13.00.00| and
\verb|maintypes = 14.00.00.00| only.) The result of this download is summarized
in the first row of Table~\ref{tab2}, whose second row is brought in directly
from Table~\ref{tab1}. Table~\ref{tab2}, therefore, summarizes the totality of
all objects used in this work.

\begin{table}[t]
\caption{Object counts in the data set used.}
\label{tab2}
\centering
\makebox[\textwidth][c]{\begin{tabular}{lccccc}
\hline
Data availability & \multicolumn{3}{c}{BPMG, cf.\ Table~\ref{tab1}} & Other & \\
\cline{2-4}
and quality & Table~\ref{bk} & Table~\ref{bn} & Table~\ref{bx} & objects & Total \\
\hline
Complete data, AAAAC & \hskip0.5em 81 & 18 & \hskip0.5em 5 & 1\,586\,324 & 1\,586\,428\\
Complete data, others & \hskip0.5em 32 & 20 & \hskip0.5em 6 & \hskip3.33em 0 & \hskip2.83em 58 \\
Complete data (total) & 113 & 38 & 11 & 1\,586\,324 & 1\,586\,486 \\
\hline
\end{tabular}
}
\end{table}

\paragraph{Data processing.}
The first processing step of all the data summarized in Table~\ref{tab2} was to
obtain $U$, $V$, and $W$ for each object. This was achieved through the usual
transformations \cite{js87}, using the J2000.0 reference system, and was
followed by a second step, comprising two runs of PCA after standardization of
the data.

For processing by PCA the data must be arranged as an $n\times p$ matrix $X$,
where $p$ is the data set's number of dimensions and $n$ is the number of
$p$-dimensional data points. The essence of PCA is the calculation of a
$p\times p$ rotation matrix $R$ from the covariance matrix of the columns of $X$
and then using it to project the $n$ data points onto $p$ new orthogonal
directions, yielding the rotated matrix $XR$. The row-$i$, column-$k$ element of
$XR$ is the so-called $k$th principal component (PC~$k$) of the $i$th data
point. Thus, in the present context of $p=4$, the standardized $U,V,W,G$ get
combined into PC~1 through PC~4. Unlike the columns of $X$, those of $XR$ are
uncorrelated with one another, and moreover PC~1 explains more of the variance
in the data than does PC~2, this one more than PC~3, etc. These two properties
justify the typical use of PCA (which is dimensionality reduction), but here we
exploit them differently: uncorrelatedness is used to afford better visual
exploration; the variance-wise primacy of PC~1, in turn, is used in the
determination of the PC-space version of the four-dimensional straight line that
seems to bind BPMG members together.

The first PCA run operated on all quality-AAAAC objects accounted for in
Table~\ref{tab2}, therefore with $n=1\,586\,428$. The resulting $R$ can be
viewed as a model, learned from high-quality data, of how to project
standardized $U,V,W,G$ points onto PC space.

The 41 quality-AAAAC objects in Tables~\ref{bk}(*),~\ref{bn}(*), and~\ref{bx}
were then taken to constitute our set of bona fide members. The 121 remaining
objects in Tables~\ref{bk},~\ref{bn}, and~\ref{bx} with complete data were taken
as candidate members. The bona fide members' projections onto PC space were
collected in a new $n\times p$ data matrix, now denoted by $X'$ and with $n=41$,
and used as the basis for the second PCA run. This second run did not aim
another rotation of the $n$ data points, but merely the determination of the
best orthogonal fit of a four-dimensional straight line to them. Letting
$\boldsymbol{\mu}=(\mu_1,\ldots,\mu_4)$ be the means vector of the $n$ points,
this was achieved by first subtracting $\boldsymbol{\mu}$ off each row of $X'$,
then running PCA to determine the new rotation matrix $R'$. Let
$\boldsymbol{a}_1$ be the first column of $R'$, that is, the vector to which the
first component (now called PC~$1'$) of the data in $X'$ refers. In parametric
form, and for $-\infty<t<\infty$, the straight line going through
$\boldsymbol{\mu}$ in the direction of $\boldsymbol{a}_1$ is
$\boldsymbol{f}(t)=\boldsymbol{\mu}+t\boldsymbol{a}_1$. Because PC~$1'$ accounts
for more variance in $X'$ than any of the other three principal components,
$\boldsymbol{f}(t)$ is the straight line to which all $n$ points are closest in
the least-squares sense and is therefore the desired four-dimensional straight
line. We henceforth view $\boldsymbol{f}(t)$ as a linear model of the BPMG and
refer to it simply as PC~$1'$.

\section{Results}
\label{res}

\begin{figure}[p]
\centering
\makebox[\textwidth][c]{\includegraphics[scale=0.50]{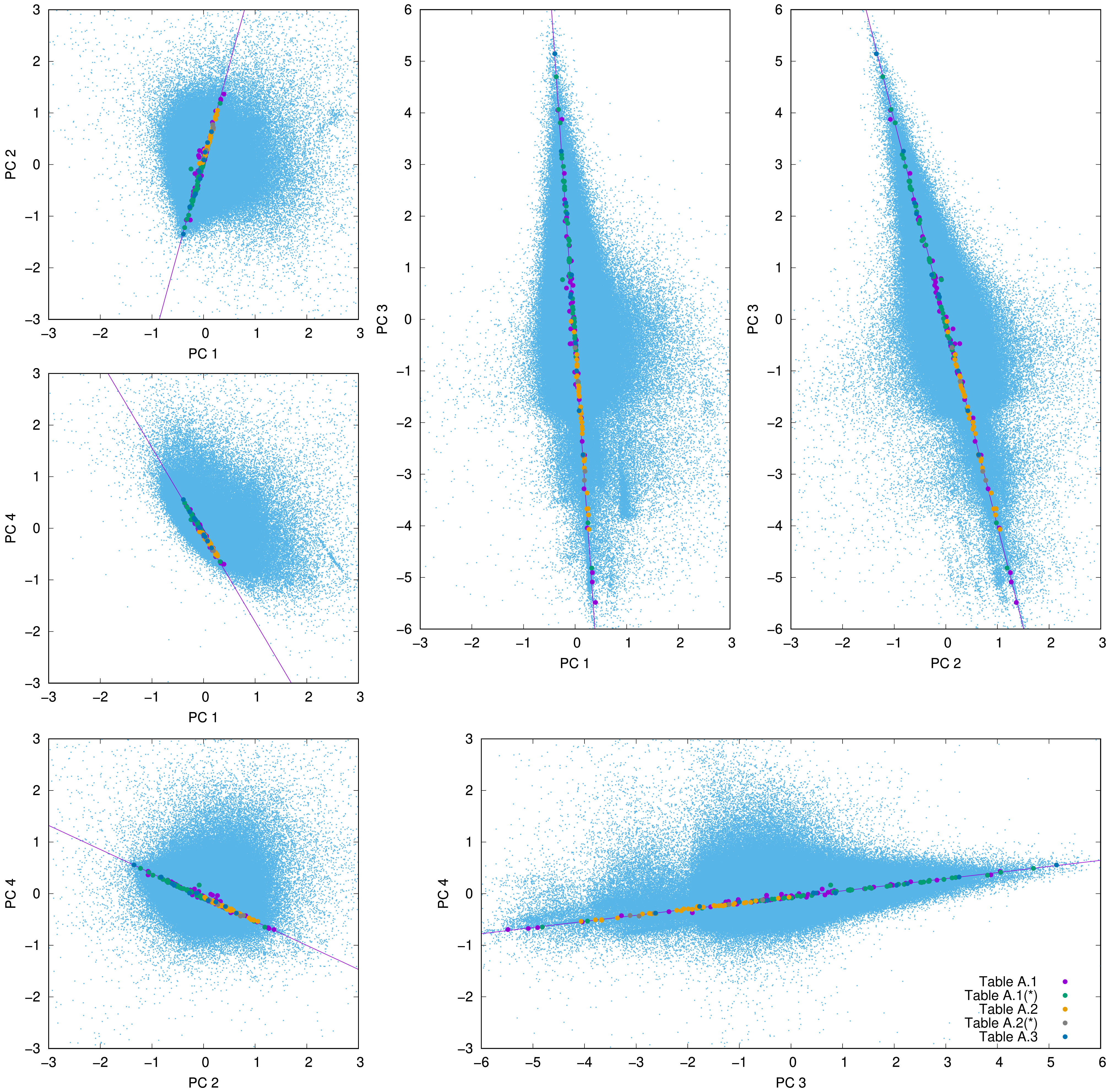}}
\caption{Two-dimensional projections of the full data set of Table~\ref{tab2}
after standardization and rotation, for all six pairs of distinct
principal-component axes. The straight line PC~$1'$ is projected as well.}
\label{fig1}
\end{figure}

Our results are summarized as the two-dimensional plots in the six panels of
Figure~\ref{fig1}. Each panel zooms in on a different two-dimensional projection
of the data set of Table~\ref{tab2} after standardization and rotation, with the
quality-AAAAC ``Other objects'' represented as tiny dots in the background and
the bona fide and candidate members as larger, color-coded dots in the
foreground. It must be noted that the latter include both the 104 quality-AAAAC
objects and the remaining 58, lower-quality ones. (One technicality is that
these 58 objects had to undergo the same standardization as the others before
rotation, being altered by means and standard deviations in which they did not
participate.) Visible in all panels is the corresponding projection of PC~$1'$
as well. 

\begin{figure}[t]
\centering
\includegraphics[scale=0.95]{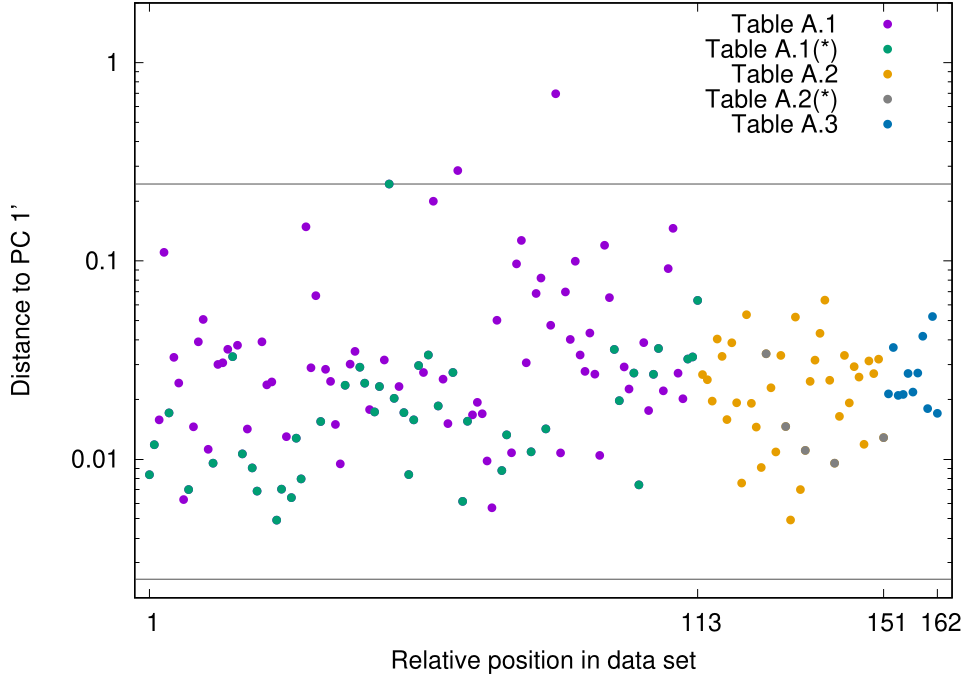}
\caption{Distances to PC~$1'$ for the BPMG objects in Table~\ref{tab2}. The
horizontal gray lines mark the lower and upper thresholds (about $0.002$ and
$0.244$) for outlier flagging.}
\label{fig2}
\end{figure}

While the plots in Figure~\ref{fig1} seem to suggest a great proximity of the
bona fide members to PC~$1'$, to varying degrees this can also be said of
the candidate members. A better view of these distances for all 162 objects of
Table~\ref{tab2} is given in Figure~\ref{fig2}. Not only does this figure show
that distances are in fact spread more widely than we might suppose by simply
examining Figure~\ref{fig1}, it tellingly also reveals that some of those
objects do indeed lie farther apart from PC~$1'$ than do the bona fide ones or
those that lie, so to speak, in their ballpark. It then makes sense that we
should try and set aside the ones that would be flagged as outliers by some
criterion. Most such criteria assume the data to follow a normal distribution,
though testing them for this is inevitably affected by the presence of the very
outliers to be eventually flagged.

\begin{figure}[t]
\centering
\includegraphics[scale=0.95]{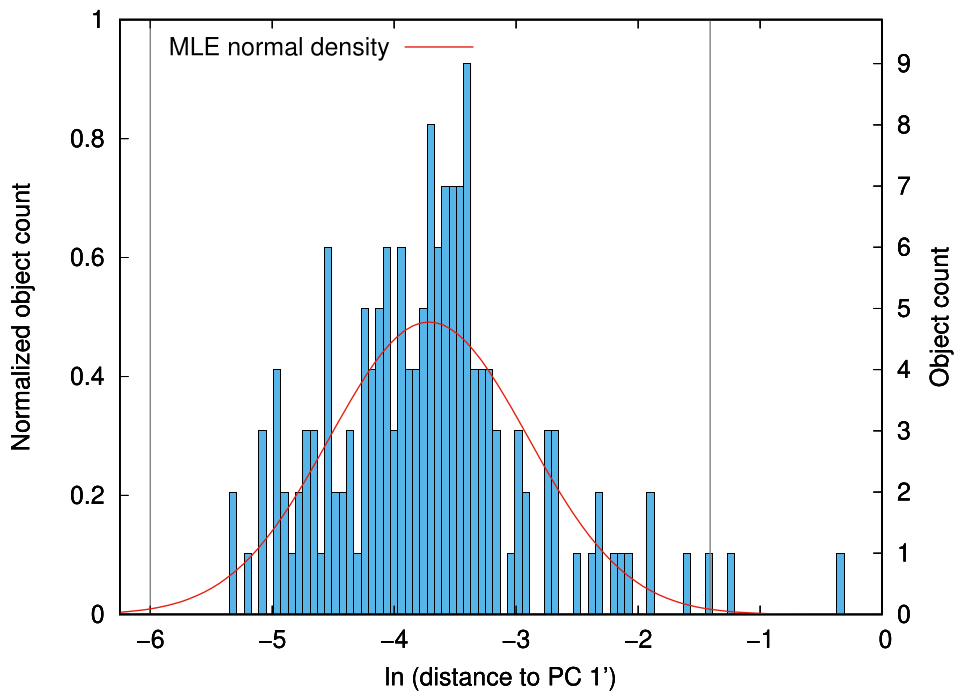}
\caption{Normalized histogram and MLE normal density of the natural logarithms
of distances to PC~$1'$ for the BPMG objects in Table~\ref{tab2}. The
corresponding original object counts are given on the vertical axis on the right
for context. The vertical gray lines mark the lower and upper thresholds (about
$-6$ and $-1.41$) for outlier flagging.}
\label{fig3}
\end{figure}

Distances to PC~$1'$ are definitely not normally distributed, but visually
inspecting the distribution of their natural logarithms does change this
significantly, as illustrated in Figure~\ref{fig3}. That is, the distances
themselves seem to follow an approximately lognormal distribution. Even this is
far from fully reliable, though, as the MLE (maximum-likelihood estimate) normal
density plotted in the same figure suggests. In cases such as this, customarily
one resorts to outlier-flagging methods that substitute the median for the mean
and the MAD (median absolute deviation) for the standard deviation. In general
this provides some added robustness in the face of uncertain normality of the
data.

A widely used method that does this is the modified Z-score method \cite{ih93}.
For $\ln(\boldsymbol{d})$ denoting the vector with the natural logarithms of all
162 distances to PC~$1'$ and $\ln(d_i)$ its $i$th element, this method's
recommendation is to flag the $i$th object as an outlier worthy of further
investigation if
\begin{equation}
\bigl\vert \ln(d_i)-\mathrm{median}(\ln(\boldsymbol{d}))\bigr\vert>
\frac
{3.5\,\mathrm{median}\bigl(\bigl\vert\ln(d_k)-
\mathrm{median}(\ln(\boldsymbol{d}))\bigr\vert\bigr)}
{0.6745}.
\end{equation}
In this expression, the range of $k$ is all of $\ln(\boldsymbol{d})$. For
$\ln(d_i)<\mathrm{median}(\ln(\boldsymbol{d}))$, the equation above leads to a
threshold of about $-6$, below which flagging is recommended (that is,
approximately for distances below $e^{-6}\approx 0.002$). For
$\ln(d_i)>\mathrm{median}(\ln(\boldsymbol{d}))$, a threshold of about $-1.41$ is
obtained, above which flagging is recommended as well (that is, approximately
for distances above $e^{-1.41}\approx 0.244$). These thresholds are marked both
as horizontal gray lines in Figure~\ref{fig2} and as vertical gray lines in
Figure~\ref{fig3}, setting apart no objects lying strictly below the lower
threshold and only two lying strictly above the upper threshold. These possible
outliers are brought in from Table~\ref{bk} and listed in Table~\ref{tab3} in
decreasing order of distance to PC~$1'$.

\begin{table}[t]
\caption{Possible outliers relative to distances to PC~$1'$.}
\label{tab3}
\centering
\makebox[\textwidth][c]{\begin{tabular}{llcccccc}
\hline
& & \multicolumn{5}{c}{Data quality} & Distance \\
\cline{3-7}
2MASS J & SIMBAD (main identifier) & $\alpha,\delta$ & $\mu_\alpha,\mu_\delta$ & $\pi$  & $\rho$ & $G$ & to PC~$1'$ \\
\hline
00325584-4405058 & EROS-MP J0032-4405		& A & A & A & B & C & 0.696 \\
20135152-2806020 & 2MASS J20135152-2806020	& A & A & A & A & C & 0.285 \\
\hline
\end{tabular}
}
\end{table}

Except for $\rho$, the currently available spatial and kinematic data for the
two possible outliers listed in Table~\ref{tab3} (and to which the quality
indicators in the table refer) all come from \emph{Gaia} EDR3. These two
objects, however, were included as known BPMG members in \cite{saklf17} from
earlier literature, therefore prior to the availability of any data from the
\emph{Gaia} mission. 2MASS J00325584-4405058 was brought in from 
\cite{gldma15}, where it is listed as a candidate member despite the absence of
a reliable $\rho$ measurement. It was moreover not tested by the proposed
selection process in \cite{saklf17}, having therefore undergone no further
confirmation. 2MASS J20135152-2806020, in turn, is from \cite{lda16}, where the
authors caution that membership confirmation depends on further verification,
``typically with radial velocity measurements.'' It passed the test in
\cite{saklf17}, based in part on $\rho=-5.81\pm 0.5\textrm{ km s}^{-1}$, which
seems to agree with the current best estimate
($\rho=-6.53\pm 0.24\textrm{ km s}^{-1}$, cf.\ \cite{frcgflfwttblr16}) only by
the thinnest margin.

It must also be noted that, though not listed as a possible outlier in
Table~\ref{tab3}, 2MASS J18141047-3247344 (HD 319139, a binary system hosting a
protoplanetary disk), the third farthest object from PC~$1'$, misses the upper
threshold by only a hair's breadth. It too was brought into \cite{saklf17} from
elsewhere \cite{tqms08} and passed the authors' selection process. However, this
seems to have taken into account $\rho=-13.3\pm 7.7\textrm{ km s}^{-1}$ while
the best estimate now available is markedly different
($\rho=-51.01\pm 0.59\textrm{ km s}^{-1}$, cf.\ \cite{mgbbboarfl20}). This
notwithstanding, this object's absence from Table~\ref{tab3} is of course amply
supported by its presence amid our bona fide members, which we recall comprises
only quality-AAAAC objects in the list from \cite{chbpgppvgs21};
cf.\ Table~\ref{bk}(*). In any event, it seems advisable to keep monitoring the
measurements of $\rho$ for this object.

\section{Concluding remarks}
\label{concl}

Deciding on an object's membership status in the BPMG (and, in general, in any
other young stellar moving group) must rely on the measurement of several
astrophysical properties that can be hard and expensive to acquire. In this
work we have introduced a new tool that can be used in the process. At the heart
of this tool is PCA, one of the staple techniques in data science, here used for
its ability to rotate variables so they become uncorrelated and to automatically
identify the one resulting variable to which more variance in the data can be
ascribed than to any other. The outcome of interest is the four-dimensional
straight line we have called PC~$1'$.

In the case of the BPMG, we demonstrated that PC~$1'$ can function as a linear
model of the group, since distances from candidate members to it can be used as
a kind of proxy for group membership. We reached this conclusion by partitioning
the membership compilation of \cite{chbpgppvgs21} into a set of bona fide
members and a set of candidate members, and then enlarging the latter set with
the inclusion of several objects from the compilation of \cite{saklf17}.
PC~$1'$ was constructed from the set of bona fide members, using further
quality-AAAAC data on more than 1.5 million objects. For the case of the BPMG
these data were ultimately standardized versions of $U,V,W,G$, but alternatives
can certainly be considered. Of course, one downside of the overall approach is
that complete data are needed for all participating objects, but again obtaining
such data is a constantly pursued goal. In general, it is a matter of time
before they become available.

\subsection*{Acknowledgments}

We thank Vladimir G. Ortega for introducing us to the topic of the BPMG and Joel
Kastner for commenting on an earlier version. We acknowledge partial support
from Conselho Nacional de Desenvolvimento Cient\'\i fico e Tecnol\'ogico (CNPq),
Coordena\c c\~ao de Aperfei\c coamento de Pessoal de N\'\i vel Superior (CAPES),
and a BBP grant from Funda\c c\~ao Carlos Chagas Filho de Amparo \`a Pesquisa do
Estado do Rio de Janeiro (FAPERJ).

\bibliography{bpic}
\bibliographystyle{unsrt}

\appendix
\setcounter{table}{0}
\renewcommand{\thetable}{\Alph{section}.\arabic{table}}

\section{Supplementary tables}

This appendix contains Tables~\ref{bk}--\ref{bx}.

\begin{table}[ht]
\caption{Identifiers and data quality of the BPMG members listed in Table~4
(first part, known members) of \cite{saklf17}.}
\label{bk}
\centering
\makebox[\textwidth][c]{\begin{tabular}{lclccccc}
\hline
& Also & & \multicolumn{5}{c}{Data quality} \\
\cline{4-8}
2MASS J & in \cite{chbpgppvgs21} & SIMBAD (main identifier) & $\alpha,\delta$ & $\mu_\alpha,\mu_\delta$ & $\pi$  & $\rho$ & $G$ \\
\hline
00065008-2306271 &*& HD 203			& A & A & A & C & C \\
00172353-6645124 &*& SCR J0017-6645		& A & A & A & A & C \\
00233468+2014282 & & V* FK Psc			& B & B & - & A & - \\
00274534-0806046 & & SCR J0027-0806		& A & A & A & B & C \\
00275023-3233060 &*& GJ 2006 A			& A & A & A & A & C \\
00275035-3233238 & & GJ 2006 B			& A & A & A & A & C \\
00325584-4405058 & & EROS-MP J0032-4405		& A & A & A & B & C \\
00440332+0228112 & & 2MASS J00440332+0228112	& C & D & - & - & - \\
00464841+0715177 & & 2MASS J00464841+0715177	& A & A & A & A & C \\
01071194-1935359 & & RX J0107.1-1935		& B & B & - & A & C \\
01112542+1526214 & & LP 467-16			& B & C & D & A & C \\
01132817-3821024 & & CD-39 325			& B & C & - & A & - \\
01294256-0823580 & & 2MASS J01294256-0823580	& A & A & A & - & C \\
01351393-0712517 &*& Barta 161 12		& A & A & A & A & C \\
01354915-0753470 & & GPM 23.953841-07.895919	& A & A & A & C & C \\
01365516-0647379 & & G 271-110			& A & A & A & A & C \\
01373545-0645375 & & V* EX Cet			& A & A & A & A & C \\
01373940+1835332 &*& BD+17 232			& B & B & - & A & - \\
01535076-1459503 & & RX J0153.5-1459		& B & C & - & A & - \\
02172472+2844305 & & HD 14082B			& A & A & A & A & C \\
02172527+2844423 &*& HD 14082			& A & A & A & A & C \\
02175601+1225266 & & PM J02179+1225		& A & A & A & A & C \\
02232663+2244069 & & LP 353-51			& A & A & A & A & C \\
02261625+0617331 & & HD 15115			& A & A & A & A & C \\
02272804+3058405 & & BD+30 397B			& A & A & A & A & C \\
02272924+3058246 &*& BD+30 397			& A & A & A & A & C \\
\hline
\end{tabular}
}
\end{table}

\addtocounter{table}{-1}
\begin{table}[ht]
\caption{Continued.}
\centering
\makebox[\textwidth][c]{\begin{tabular}{lclccccc}
\hline
& Also & & \multicolumn{5}{c}{Data quality} \\
\cline{4-8}
2MASS J & in \cite{chbpgppvgs21} & SIMBAD (main identifier) & $\alpha,\delta$ & $\mu_\alpha,\mu_\delta$ & $\pi$  & $\rho$ & $G$ \\
\hline
02282694+0218331 & & 2MASS J02282694+0218331	& A & A & A & - & C \\
02303239-4342232 & & CD-44 753			& A & A & A & A & C \\
02304623-4343493 & & UCAC2 13050114		& A & A & A & A & C \\
02365171-5203036 & & EXO 0235.2-5216		& A & A & A & B & C \\
02412589+0559181 & & BD+05 378			& A & A & A & A & C \\
02501167-0151295 & & TVLM 831-154910		& A & A & A & - & C \\
02534448-7959133 & & SIPS J0253-7959		& A & A & A & - & C \\
03111547+0106307 & & [BHR2005] 832-2		& A & A & A & D & C \\
03323578+2843554 & & RX J0332.6+2843		& B & C & - & A & C \\
03350208+2342356 &*& 2MASSW J0335020+234235	& A & A & A & B & C \\
04373613-0228248 &*& * 51 Eri			& A & A & A & A & C \\
04373746-0229282 & & StKM 1-497			& A & A & A & A & C \\
04433761+0002051 &*& 2MASSI J0443376+000205	& A & A & A & A & C \\
04435686+3723033 & & V* V962 Per		& A & A & A & A & C \\
04480085+1439583 & & UCAC2 36944937		& A & A & A & - & C \\
04480258+1439516 & & UCAC3 210-20333		& A & A & A & - & C \\
04593483+0147007 &*& V* V1005 Ori		& A & A & A & A & C \\
05004714-5715255 &*& CD-57 1054			& A & A & A & A & C \\
05015881+0958587 &*& LP 476-207			& A & A & A & B & - \\
05082729-2101444 & & UCAC4 345-006842		& A & A & A & A & C \\
05120636-2949540 & & 2MASS J05120636-2949540	& C & C & - & - & C \\
05195327+0617258 & & GSC2.3 N9OB003170		& A & A & A & - & C \\
05200029+0613036 & & RX J0520.0+0612		& A & A & A & A & C \\
05203182+0616115 & & RX J0520.5+0616		& A & A & A & A & C \\
05241914-1601153 & & PM J05243-1601		& B & C & - & A & - \\
05270477-1154033 &*& V* AF Lep			& A & A & A & A & C \\
05294468-3239141 &*& RX J0529.7-3239		& A & A & A & A & C \\
05335981-0221325 & & RX J0534.0-0221		& A & A & A & A & C \\
05471708-5103594 &*& * bet Pic			& A & A & A & A & C \\
06131330-2742054 &*& RX J0613.2-2742		& A & A & A & A & C \\
06182824-7202416 &*& V* AO Men			& A & A & A & A & C \\
08173943-8243298 & & 1RXS J081742.4-824331	& A & A & A & B & C \\
08224744-5726530 & & L 186-67			& A & A & A & A & C \\
09360429+3733104 & & HD 82939			& A & A & A & A & C \\
09361593+3731456 & & MCC 549			& A & A & A & D & C \\
10141918+2104297 & & MCC 124			& A & A & A & A & C \\
10172689-5354265 &*& TWA 22			& A & A & A & A & C \\
10593834+2526155 & & HD 95174A			& A & A & A & A & C \\
10593870+2526138 & & HD 95174B			& A & A & A & D & C \\
11493184-7851011 & & V* DZ Cha			& A & A & A & A & C \\
\hline
\end{tabular}
}
\end{table}

\addtocounter{table}{-1}
\begin{table}[ht]
\caption{Continued.}
\centering
\makebox[\textwidth][c]{\begin{tabular}{lclccccc}
\hline
& Also & & \multicolumn{5}{c}{Data quality} \\
\cline{4-8}
2MASS J & in \cite{chbpgppvgs21} & SIMBAD (main identifier) & $\alpha,\delta$ & $\mu_\alpha,\mu_\delta$ & $\pi$  & $\rho$ & $G$ \\
\hline
13545390-7121476 & & UPM J1354-7121		& A & A & A & A & C \\
14141700-1521125 & & L 836-122			& A & A & A & - & C \\
14142141-1521215 & & HD 124498			& A & A & A & A & - \\
14252913-4113323 & & SCR J1425-4113AB		& A & A & A & B & C \\
15385679-5742190 & & HD 139084B			& A & A & A & A & C \\
15385757-5742273 &*& HD 139084			& A & A & A & A & C \\
16181789-2836502 & & * d Sco			& A & A & A & A & C \\
16430128-1754274 & & ASAS J164301-1754.4	& A & A & A & A & C \\
16572029-5343316 &*& TYC 8726-1327-1		& A & A & A & A & C \\
17150219-3333398 & & 1RXS J171502.4-333344	& A & A & A & C & C \\
17150362-2749397 & & CD-27 11535		& A & A & A & C & C \\
17172550-6657039 &*& HD 155555			& A & A & A & A & C \\
17173128-6657055 & & HD 155555C			& A & A & A & A & C \\
17292067-5014529 & & GSC 08350-01924		& C & D & - & B & - \\
17295506-5415487 &*& CD-54 7336			& A & A & A & A & C \\
17414903-5043279 &*& HD 160305			& A & A & A & C & C \\
17483374-5306433 &*& HD 161460			& A & A & A & C & C \\
18030341-5138564 &*& HD 164249			& A & A & A & A & C \\
18030409-5138561 & & HD 164249B			& A & A & A & A & C \\
18064990-4325297 &*& CD-43 12272		& A & A & A & A & - \\
18141047-3247344 &*& HD 319139			& A & A & A & A & C \\
18142207-3246100 & & ASAS J181422-3246.2	& A & A & A & C & C \\
18151564-4927472 & & 2MASS J18151564-4927472	& A & A & A & C & C \\
18195221-2916327 &*& HD 168210			& A & A & A & C & C \\
18202275-1011131 & & BD-10 4662			& A & A & A & A & - \\
18420483-5554126 &*& UCAC4 171-199133		& A & A & A & A & C \\
18420694-5554254 & & 1RXS J184206.5-555426	& A & A & A & A & C \\
18452691-6452165 &*& HD 172555			& A & A & A & C & C \\
18453704-6451460 & & CD-64 1208			& A & A & A & C & C \\
18465255-6210366 & & Smethells 20		& A & A & A & C & C \\
18480637-6213470 &*& HD 173167			& A & A & A & A & C \\
18504448-3147472 &*& CD-31 16041		& A & A & A & C & C \\
18530587-5010499 &*& V* PZ Tel			& A & A & A & A & C \\
18580415-2953045 &*& 1RXS J185803.4-295318	& A & A & A & C & C \\
18580464-2953320 & & UCAC2 19527490		& C & B & - & A & C \\
19102820-2319486 & & 1SWASP J191028.18-231948.0	& C & - & - & A & C \\
19114467-2604085 & & CD-26 13904		& A & A & A & C & C \\
19225122-5425263 & & * eta Tel			& A & A & A & C & C \\
19225894-5432170 &*& HD 181327			& A & A & A & A & C \\
19233820-4606316 &*& ASAS J192338-4606.4	& A & A & A & A & C \\
\hline
\end{tabular}
}
\end{table}

\addtocounter{table}{-1}
\begin{table}[ht]
\caption{Continued.}
\centering
\makebox[\textwidth][c]{\begin{tabular}{lclccccc}
\hline
& Also & & \multicolumn{5}{c}{Data quality} \\
\cline{4-8}
2MASS J & in \cite{chbpgppvgs21} & SIMBAD (main identifier) & $\alpha,\delta$ & $\mu_\alpha,\mu_\delta$ & $\pi$  & $\rho$ & $G$ \\
\hline
19243494-3442392 & & 2MASS J19243494-3442392	& A & A & A & A & C \\
19312434-2134226 & & 2MASS J19312434-2134226	& A & A & A & B & C \\
19355595-2846343 &*& 2MASS J19355595-2846343	& A & A & A & A & C \\
19560294-3207186 & & UCAC3 116-474938		& A & A & A & A & C \\
19560438-3207376 &*& PM J19560-3207		& A & A & A & A & C \\
20004841-7523070 & & SIPS J2000-7523		& A & A & A & B & C \\
20013718-3313139 & & UCAC4 284-205440		& A & A & A & A & C \\
20041810-2619461 & & HD 190102			& A & A & A & A & C \\
20055640-3216591 & & V* V5663 Sgr		& A & A & A & B & C \\
20090521-2613265 &*& HD 191089			& A & A & A & A & C \\
20100002-2801410 & & SCR J2010-2801		& B & C & - & A & C \\
20135152-2806020 & & 2MASS J20135152-2806020	& A & A & A & A & C \\
20333759-2556521 &*& SCR J2033-2556		& A & A & A & A & C \\
20334670-3733443 & & SIPS J2033-3733		& A & A & A & - & C \\
20415111-3226073 & & V* AT Mic			& A & A & A & A & - \\
20434114-2433534 &*& EM* StHA 182		& B & C & - & A & - \\
20450949-3120266 &*& V* AU Mic			& A & A & A & A & C \\
20554767-1706509 &*& HD 199143			& A & A & A & C & C \\
20560274-1710538 & & HD 358623			& B & C & D & A & - \\
21100461-1920302 &*& UCAC4 354-189365		& A & A & A & B & C \\
21100535-1919573 & & BPS CS 22898-0065		& A & A & A & A & C \\
21103096-2710513 & & 2MASS J21103096-2710513	& A & A & A & A & C \\
21103147-2710578 & & ** BRG 32A			& A & A & A & A & C \\
21140802-2251358 & & 2MASS J21140802-2251358	& C & C & C & - & - \\
21212446-6654573 & & V* V390 Pav		& A & A & A & E & C \\
21212873-6655063 & & PM J21214-6655		& A & A & A & A & C \\
21374019+0137137 & & RX J2137.6+0137		& A & A & A & A & C \\
22004158+2715135 & & RX J2200.7+2715		& A & A & A & A & C \\
22081363+2921215 & & 2MASSW J2208136+292121	& C & C & C & B & - \\
22424896-7142211 &*& CPD-72 2713		& A & A & A & A & C \\
22445794-3315015 &*& V* WW PsA			& A & A & A & A & C \\
22450004-3315258 & & V* TX PsA			& A & A & A & A & C \\
23172807+1936469 & & G 68-5			& A & A & A & A & C \\
23301341-2023271 & & G 273-59			& A & A & A & A & C \\
23314492-0244395 & & V* AF Psc			& A & A & A & A & C \\
23323085-1215513 &*& BPM 82931			& A & A & A & A & C \\
23353085-1908389 & & 2MASS J23353085-1908389	& A & A & A & - & C \\
23500639+2659519 & & RX J2350.0+2659		& A & A & A & A & C \\
23512227+2344207 & & G 68-46			& A & A & A & A & C \\
23542220-0811289 & & 2MASS J23542220-0811289	& C & D & - & - & - \\
\hline
\end{tabular}
}
\end{table}

\begin{table}[ht]
\caption{Identifiers and data quality of the BPMG members listed in Table~4
(second part, new members) of \cite{saklf17}.}
\label{bn}
\centering
\makebox[\textwidth][c]{\begin{tabular}{lclccccc}
\hline
& Also & & \multicolumn{5}{c}{Data quality} \\
\cline{4-8}
2MASS J & in \cite{chbpgppvgs21} & SIMBAD (main identifier) & $\alpha,\delta$ & $\mu_\alpha,\mu_\delta$ & $\pi$  & $\rho$ & $G$ \\
\hline
00164976+4515417 & & 2MASS J00164976+4515417	& C & E & - & A & - \\
00193931+1951050 & & 2MASS J00193931+1951050	& A & A & A & A & C \\
00194303+1951117 & & RX J0019.7+1951		& A & A & A & A & C \\
00281434-3227556 & & GR* 9			& A & A & A & C & C \\
00413538-5621127 & & DENIS J004135.3-562112	& A & A & A & B & C \\
00482667-1847204 & & 2MASS J00482667-1847204	& A & A & A & A & C \\
00501752+0837341 & & PM J00502+0837		& A & A & A & B & C \\
01303534+2008393 & & 2MASS J01303534+2008393	& A & A & A & A & C \\
02241739+2031513 &*& 2MASS J02241739+2031513	& A & A & A & B & C \\
02335984-1811525 & & RX J0234.0-1811		& B & C & - & A & - \\
02450826-0708120 & & 2MASS J02450826-0708120	& A & A & A & C & C \\
02485260-3404246 & & UCAC3 112-6119		& A & A & A & A & C \\
02495639-0557352 & & 2MASS J02495639-0557352	& A & A & A & A & C \\
03255277-3601161 & & UCAC2 16305530		& A & A & A & C & C \\
03363144-2619578 & & SCR J0336-2619		& A & A & A & C & C \\
03370343-3042318 & & 2MASS J03370343-3042318	& A & A & A & C & C \\
03393700+4531160 & & PM J03396+4531		& A & A & A & B & C \\
03550477-1032415 & & 2MASS J03550477-1032415	& A & A & A & A & C \\
04232720+1115174 & & 2MASS J04232720+1115174	& A & A & A & A & C \\
05015665+0108429 & & PM J05019+0108		& A & A & A & A & C \\
05061292+0439272 &*& RX J0506.2+0439		& A & A & A & C & C \\
05363846+1117487 & & PM J05366+1117		& A & A & A & A & C \\
13215631-1052098 & & PM J13219-1052		& A & A & A & C & C \\
15063505-3639297 & & 2MASS J15063505-3639297	& A & A & A & B & C \\
18011345+0948379 & & UPM J1801+0948		& A & A & A & B & C \\
18055491-5704307 & & UPM J1805-5704		& A & A & A & A & C \\
18090694-7613239 & & SIPS J1809-7613		& A & A & A & A & C \\
18092970-5430532 & & 2MASS J18092970-5430532	& A & A & A & A & C \\
18435838-3559096 & & 2MASS J18435838-3559096	& A & A & A & C & C \\
18471351-2808558 & & 2MASS J18471351-2808558	& A & A & A & B & C \\
19082195-1603249 &*& 2MASS J19082195-1603249	& A & A & A & A & C \\
19260075-5331269 & & 2MASS J19260075-5331269	& A & A & A & C & C \\
19300396-2939322 & & 2MASS J19300396-2939322	& A & A & A & B & C \\
20083784-2545256 & & 2MASS J20083784-2545256	& A & A & A & B & C \\
20085368-3519486 & & UCAC4 274-196167		& B & C & - & A & - \\
21200779-1645475 & & 2MASS J21200779-1645475	& A & A & A & A & C \\
21384755+0504518 & & 2MASS J21384755+0504518	& A & A & A & B & C \\
22085034+1144131 &*& PM J22088+1144		& A & A & A & B & C \\
22334687-2950101 & & 2MASS J22334687-2950101	& A & A & A & A & C \\
23010610+4002360 & & 2MASS J23010610+4002360	& A & A & A & A & C \\
23355015-3401477 &*& SIPS J2335-3401		& A & A & A & A & C \\
\hline
\end{tabular}
}
\end{table}

\begin{table}[ht]
\caption{Identifiers and data quality of the BPMG members listed in Table~A1 of
\cite{chbpgppvgs21}, excluding those already appearing in Table~\ref{bk} or
Table~\ref{bn}, where they are marked with asterisks.}
\label{bx}
\centering
\makebox[\textwidth][c]{\begin{tabular}{llccccc}
\hline
& & \multicolumn{5}{c}{Data quality} \\
\cline{3-7}
2MASS J & SIMBAD (main identifier) & $\alpha,\delta$ & $\mu_\alpha,\mu_\delta$ & $\pi$  & $\rho$ & $G$ \\
\hline
05004928+1527006 & HD 286264 		& A & A & A & A & C \\
05315786-0303367 & PM J05319-0303W	& A & A & A & B & C \\
05320450-0305291 & V* V1311 Ori 	& A & A & A & A & C \\
05320596-0301159 & ESO-HA 737		& A & A & A & C & C \\
14423039-6458305 & * alf Cir 		& A & A & A & A & C \\
17453733-2824269 & HD 161247		& A & A & A & C & C \\
17520173-2357571 & PM J17520-2357	& A & A & A & B & C \\
18183181-3503026 & HD 167847B		& A & A & A & C & C \\
20524162-5316243 & HD 198472		& A & A & A & A & C \\
21354554-4218343 & [HB88] M11		& A & A & A & B & C \\
22311828-0633183 & HD 213429		& A & A & A & A & C \\
\hline
\end{tabular}
}
\end{table}

\end{document}